# Evaluating soft tissue simulation in maxillofacial surgery using pre and post-operative CT scan

M. Chabanas [a,*], Ch. Marécaux [a,b], F. Chouly [a], F. Boutault [b], Y. Payan [a]

[a] *TIMC-IMAG, UMR CNRS 5525, Université Joseph Fourier, Grenoble, France*
[b] *Service de Chirurgie Maxillo Faciale et Plastique de la Face, Hôpital Purpan, Toulouse, France*

**Abstract.** One of the most important issue in soft tissue modeling is to assess the quality of the simulations. A validation protocol is presented based on two CT scans of the patient acquired before and after cranio-maxillofacial surgery. The actual bones repositioning realized during the intervention are accurately measured and reproduced. A evaluation of the soft tissue deformation is then computed using a finite element model of the face. The simulations are therefore compared, qualitatively and quantitatively, with the actual outcome of the surgery. This protocol enable to rigorously evaluate different modeling methods, and to assess the clinical relevance of soft tissue simulation in maxillofacial surgery.

## 1. Introduction

Modeling the human soft tissue is of growing interest in medical and computer science fields, with a wide range of applications such as physiological analysis, surgery planning, or interactive simulation for training purpose [1]. In maxillofacial surgery, the correction of face dismorphosis is addressed by surgical repositioning of bone segments, e.g. the mandible, maxilla or zygomatic bone. The aim is to correct the skull deformities and the dental occlusion, and to improve the aesthetic of the patient face. A model of the patient face to simulate the morphological modifications following bone repositioning could greatly improve the planning of the intervention, for both the surgeon and the patient. Different models of the face soft tissue were proposed in the literature. A review can be found in [2].

One of the most important issue in soft tissue modeling is to assess the quality of the simulations. From a modeling point of view, it enables to evaluate different methods, for example a linear elastic versus a non-linear model. This is above all essential for the surgeon since the use of a soft tissue model in actual surgical practice cannot be considered without an extensive clinical validation.

---

[*] Corresponding author. *E-mail address*: …@address.com.

Due to the extreme complexity of the tissue behavior and the medical interactions (physiological phenomenon, surgical or therapeutic intervention), there is obviously no theoretical solutions to the modeling problems. Simulations must then be compared with validation data sets which provide adequate information to assess the ground truth [3]. Although experimental systems were developed to measure the deformations of known materials under controlled constraints [4,5], there is a an important lack of such data, and neither gold nor bronze standard are available so far.

For computer-aided surgical planning, the most straightforward validation approach is to work on pre and post-operative data on real clinical cases. This paper presents a rigorous validation procedure for maxillofacial applications, based on a two CT scans acquired before and after the surgery. They provide us with the initial state of the system and enable to accurately measure the realized interaction. This gesture is then reproduced with the face soft tissue model, and the simulations are finally compared, qualitatively and quantitatively, with the actual post-operative outcomes.

## 2. Materials and methods

Few authors have proposed extended validation procedures for soft tissue modeling. In maxillofacial surgery, most of them compare their simulations with facial and profile pictures of the patient. While a qualitative comparison is always required, this method is quite inaccurate and does not afford a real tri-dimensional evaluation. The main other approach rely on the acquisition of the post-surgical patient morphology with an optical laser scanner, which enable a 3D quantitative comparison [6]. However, it is very sensitive to the accuracy of the skin surface and the registration procedure to express it in the pre-operative patient referential. Moreover, there is always an important error between the simulated intervention and the bone repositioning actually realized during the surgery. The most advanced quantitative evaluation was recently proposed by [7], who measure the distances between their simulations and a post-operative CT scan.

The evaluation protocol we propose also requires the acquisition of a pre and a post-operative. It consists in four steps:
 - measuring the bone repositioning actually realized during the surgery, by direct comparison of the pre- and post-operative data;
 - simulating the bone osteotomies and applying the measured displacements to the bone segments;
 - simulating the resulting soft tissue deformation using the biomechanical model;
 - evaluating the differences between the simulation and the post-operative data, both qualitatively and quantitatively.

*2.1 Validation data sets : pre and post-operative CT scans*

While a pre-operative CT scan is regularly acquired for complex maxillofacial procedures, the requirement of a post-operative scan could be the main drawback of our validation protocol. Indeed, it is not clinically required and represents an added invasivity in terms of radiations. However, it is clearly the data that fit the most to assess the quality of the numerical simulations. With the improvement of modern scanners and the steady reduction of irradiation, its use appears acceptable in a research context.

*2.2 Measuring the actual bones repositioning*

Since both CT scans are expressed in different coordinate systems, the first problem is to transfer the reconstructed 3D models in a same referential. These models can then be compared to measure the displacements applied to bone segments during the surgery.

These two steps are realized using mathematical tools initially developed for a 3D cephalometry project [8]. A geometrical model of the facial skeleton is built out of anatomical landmarks that are defined in the CT slices. Landmarks in areas that are not modified during the surgery are used to express the pre and post-operative data in a common referential. Then, landmarks located on each bone segment (e.g. the mandible and maxilla) are determined in both datasets. Using the Arun method [9], the displacements actually applied during the surgery can therefore be computed as a rotation and a translation (Fig. 1).

Although the anatomical landmarks are manually positioned on the CT slices, it has been shown that their repeatability is always less than .5 mm, which yields to very acceptable results in the displacements measurements. To further improve the accuracy of the measurement, a rigid registration can be carried out to match the 3D geometric reconstructions of the bone segments. This is mostly useful in areas where no anatomical landmark clearly appears, like the tip of the chin.

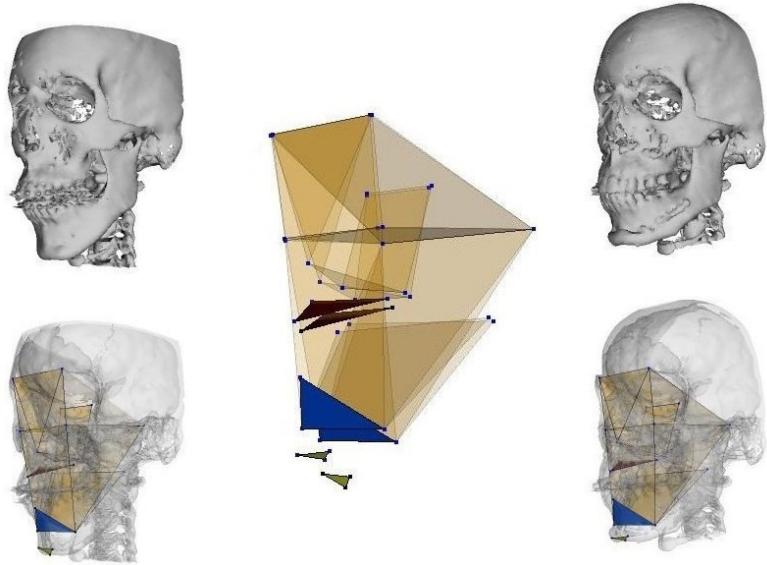

Fig. 1. Measurement of the bones repositioning realized during the surgery. Anatomical landmarks are defined in the pre- and post-operative 3D skeleton model. Points in the upper part of the skull are used to register both models in a same coordinate system. The displacement applied to a bone segment is computed by comparing a set of landmarks in the two models, then refined using rigid registration. In this example, the maxillae, the mandible and the tip of the chin (genioplasty) were repositioned.

*2.3 Simulating the surgery*

The measured surgical procedure must be reproduced on the virtual model of the patient.

Each osteotomy is simulated using a set of cutting planes interactively positioned on the 3D skeleton model. Other methods exist for a more precise definition of the osteotomy lines, using free-hand drawing on 3D polygonal surfaces [10]. This is mostly useful in case of complex surgical procedures. However, cutting planes is a more straightforward and easy to use technique, which provides fairly good results for the simulation of standard maxilla, mandible and genial osteotomies. In a normal planning process, the osteotomies are defined according to standard surgical practice. For the clinical validation, the simulation of the osteotomies are interactively checked with the post-operative 3D reconstruction, where the actual lines of cut can be clearly identified.

Once every osteotomy has been realized, each bone segment is repositioned using the actual displacements measured between the pre- and post-operative data (Fig. 2).

The deformations of the soft tissues are then simulated using a finite element model of the patient face [2]. The measured displacements define the boundary conditions for the model: inner nodes in contact with the non-modified skeleton surface are fixed, while the measured displacements are applied to the nodes on the osteotomized bone segments. Nodes around the osteotomy line are not constrained, to account for the bone-tissue separation due to the surgical access. Rest of the nodes, in the outer part of the mesh or in the mouth and cheeks area are let free to move.

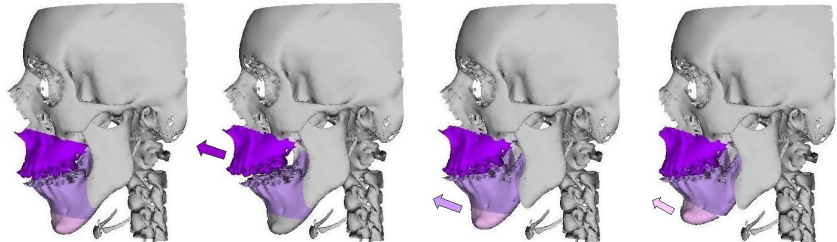

Fig. 2. Simulation of maxilla, mandible and genial osteotomies. The osteotomies are simulated using a set of planes cutting the 3D reconstruction of the patient facial skeleton. Each bone segment is then repositioned using the actual displacements measured between the pre- and post-operative data.

*2.4 Assessing the quality of the simulation: comparison with the post-operative data*

Once the outcome of the surgery has been simulated, it can be compared with the post-operative skin surface of the patient, reconstructed from the CT scan. This comparison is performed in two steps: a qualitative evaluation based on precise aesthetic and morphological criteria, followed by a quantitative measurement of the errors between the simulated and actual patient morphology.

The *qualitative* procedure consists in a visual comparison between the deformed model and the 3D reconstruction of the patient skin surface (Fig. 3). For each simulation, images are printed with different angles of view: frontal, left and right profile, left and right oblique views, upper and lower views. An interactive 3D visualization is also available to magnify some areas or to use specific angles of view. The deformed model and the patient reconstruction are observed one next to each other, then in superposition. Emphasis is given to the perception of the model quality in the most relevant morphological areas in the face: cheeks bones, lips area, chin and mandible angles.

The *quantitative* comparison consists in a distance map between the deformed model and the actual patient data (Fig. 4). It is computed using the MESH software [11], which was improved to calculate signed Euclidian distances.

## 3. Results

The evaluation procedure has been successfully applied to three clinical cases. Figures 1 to 4 show the different steps of the protocol in a case of bimaxillary repositioning along with a genioplasty. For all patients, the quantitative measurements shows errors between 1 and 1.5 mm in mean, with maximal values of 3 to 6 mm.

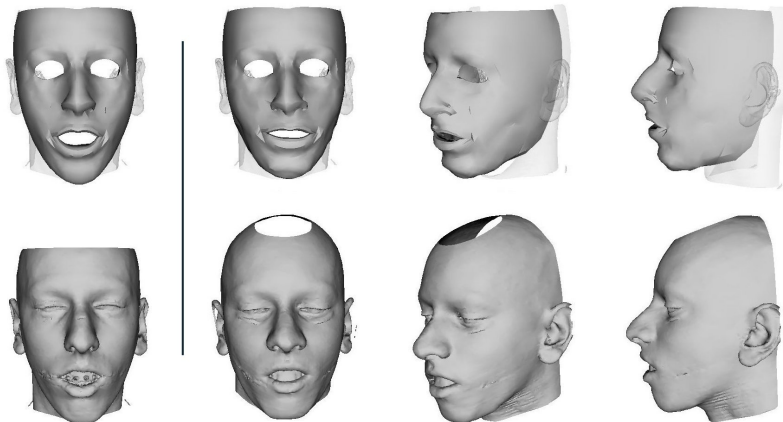

Fig. 3. Qualitative evaluation. The simulations (top) are visually compared with the 3D reconstruction of the post-operative patient skin surface (bottom). Emphasis is given to the perception of the model quality in the most relevant morphological areas in the face: cheeks bones, lips area, chin and mandible angles.

## 4. Discussion

While the numerical results first appear quite satisfying, the protocol based on qualitative and quantitative evaluation clearly emphasizes the errors in our simulations, mostly in the lips area, and the limits of our current modeling (linear elasticity, problem of contacts between the lips and the teeth). These problems will therefore be specifically addressed. The evaluation procedure has also proven to be extremely relevant to evaluate the different modeling improvements that are currently tested, like using non-linear elasticity or an hyperelastic finite element model.

Although our validation methodology appears quite complete, it still presents several limits. First, it requires a post-operative CT scan. Second, the distance map errors, which should be carefully analyzed. They are calculated as the minimal Euclidian distance between points of the model and the patient skin surface, which does not mean the distances to their "real" corresponding points. These numerical values are thus always a minimization of the true errors. Finally, errors in the neck area are difficult to estimate if the patient is not in the same posture during the two CT scans.

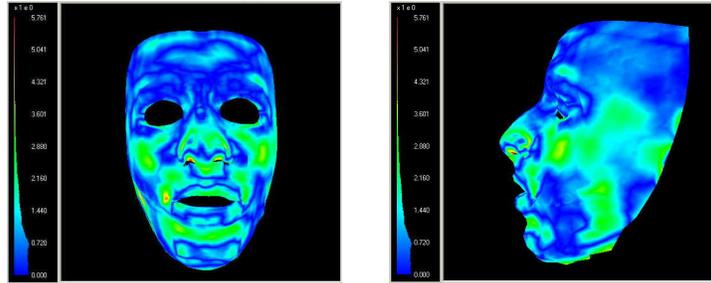

Fig. 4. Quantitative evaluation, through a distance map between the simulation and the post-operative data.

## 5. Conclusion

An extended procedure has been proposed to assess the quality of soft tissue simulation in maxillofacial surgery. Using pre and post-operative CT scans, the actual bone repositioning realized during the surgery are accurately measured. A qualitative and quantitative evaluation of the error committed during the simulation is then provided. This method has given significant results on three clinical cases.